\documentclass[aps,prl,twocolumn,superscriptaddress,amsmath,amssymb]{revtex4-1}

\usepackage[colorlinks,pdfusetitle,urlcolor=blue,citecolor=blue,linkcolor=blue,bookmarksnumbered,plainpages=false]{hyperref}

\usepackage{color}

\usepackage{bm}
\usepackage{nicefrac}
\usepackage{siunitx}

\usepackage{graphicx}
\usepackage{braket}

\newcommand{\B}[1]{\mathbf{#1}}

\newcommand{\im}{\operatorname{Im}}
\newcommand{\EpsRatioStatic}{\eta(0)}
\newcommand{\EpsRatio}{\eta}

\begin{document}

\title{Distinguishing models of surface response through the self-energy of an electron}

\author{Robert Bennett}
\affiliation{Physikalisches Institut, Albert-Ludwigs-Universit\"at Freiburg,\\ Hermann-Herder-Str. 3, D-79104 Freiburg i. Br., Germany}

\author{Stefan Yoshi Buhmann}
\affiliation{Physikalisches Institut, Albert-Ludwigs-Universit\"at Freiburg,\\ Hermann-Herder-Str. 3, D-79104 Freiburg i. Br., Germany}

\author{Claudia Eberlein}
\affiliation{Department of Physics \& Astronomy, University of Sussex, Falmer, Brighton BN1 9QH, UK}

\date{\today}

\begin{abstract} The self-energy of an electron confined between parallel surfaces with arbitrary dielectric properties is calculated. The mechanism for this effect is the surface-induced modification of the fluctuating quantised vacuum field to which the electron is coupled, thereby endowing it with a surface-dependent self-energy in broad analogy to the Casimir-Polder effect for an atom. We derive a general formula for this self-energy shift and find that its sign is different for two commonly-used models of surface response, namely the plasma model and the Drude model. We propose an experiment which could detect this difference in sign, shedding light on continuing uncertainty about the correct description of the interaction of low-frequency vacuum photons with media.  \end{abstract}

\maketitle

Quantum electrodynamics is an extremely successful description of the interaction between charges and electromagnetic fields. One of its most remarkable predictions is the existence of a fluctuating ground state, variously known as the zero-point energy or vacuum field. Often cited as a evidence that vacuum fluctuations are `real' is the attractive force between parallel plates arising from the imposition of boundary conditions on the vacuum field, this is the famous Casimir effect \cite{Casimir1948}.  Beginning with the pioneering work of Sparnaay in 1957 \cite{SPARNAAY1957}, there have been a string of experiments measuring the Casimir force in various situations \cite{Lamoreaux1997, *Lamoreaux1998, Mohideen1998, Harber2005, Decca2007}. An unexpected result was seen in \cite{Decca2007}, where it was found that theory and experiment agree if a lossless plasma model is used for the surfaces, and the apparently more realistic Drude model of a dissipative surface makes predictions inconsistent with experiment. Even more curiously, a later experiment \cite{Sushkov2011} produced the \emph{opposite} conclusion --- its results fit with the Drude model and not the undamped plasma. This has led to considerable amount of discussion over the last decade or so \cite{Bostrom2000,*Svetovoy2005,*Brevik2006,*Kim2008,*Decca2009b,*Kim2009,*bordag2009advances,*Guerout2014,*Klimchitskaya2016,*Guerout2016,*Bimonte2016,*Cherroret2017,*Hartmann2017}. This was fuelled in part by this problem's status as a dominant error in experiments aiming to probe physics beyond the standard model.  Perhaps even more importantly, the Drude-plasma question has implications for the fundamental physical question of whether virtual photons are subject to dissipation. 

Bearing this in mind, it is natural to wonder whether there any independent checks in surface-dependent vacuum QED for which theory predicts strongly differing results depending on wether a Drude or plasma model is used. Previously it has been shown that a single electron interacting with an infinite half-space is a rich test-bed for different models of surface-dependent effects due to its extreme sensitivity to the low-frequency response of the medium \cite{Eberlein2004,Bennett2012}, which can drastically change depending upon the choice of model. Here we will first use the formalism of macroscopic QED \cite{Gruner1995} to generalise this result to arbitrary geometries, and then consider a specific case for which we also propose an experiment that could conclusively determine whether Drude or plasma response is appropriate for a given material. 

We begin by considering a single electron of momentum $\hat{\B{p}}$, minimally coupled to the electromagnetic field $\{\Phi, \hat{\B{A}} \}$, so that the interaction Hamiltonian is $H_\text{int}= - \frac{e}{m} \hat{\B{p}}\cdot \hat{\B{A}} + e \Phi $. In macroscopic QED, the components of the vector potential at position $\B{r}$ in a system composed of material bodies of permittivity $\varepsilon(\B{r},\omega)$ may be expressed in terms of the bosonic operators $\hat{{a}}_{j}^\dagger(\B{r},\omega)$ and $\hat{{a}}_{j}(\B{r},\omega)$ respectively create or destroy one excitation of the combined matter-field system via \footnote{We use natural units $\hbar = c = \epsilon_0 = 1$ throughout unless otherwise specified}
\begin{align}
\hat{{A}}_i(\B{r}) =\frac{1}{\sqrt{\pi}} \int_0^\infty & d \omega \, \omega  \int d^3\B{r}' \sqrt{\im \varepsilon(\B{r}',\omega)}\notag  \\
&\times \B{G}_{ij}(\B{r},\B{r}',\omega) \hat{{a}}_{j}(\B{r}',\omega)+ \text{H.c.} \;  \label{VecPot}
\end{align}
where all geometric properties are encoded via the (classical) electromagnetic Green's function $\B{G}$ satisfying
\begin{align}
\nabla \times &  \nabla \times \B{G}(\B{r},\B{r}',\omega)  -\omega^2 \varepsilon(\B{r},\omega)  \B{G}(\B{r},\B{r}',\omega) = \mathbb{I}_3 \delta(\B{r}-\B{r}'),
\end{align}
Labelling a state with momentum eigenstate $\B{p}$ and $N$ matter-field excitations by $\ket{\B{p};N}$, and postponing discussion of the static contribution from $\Phi$, one has for the lowest-order momentum-dependent energy shift of the vacuum state 
\begin{equation} \label{VacuumEnergyShift}
\Delta E = \frac{e^2}{m^2}\sum_{\B{p}'} \frac{|\bra{\B{p}';1}\hat{\B{p}}\cdot \hat{ \mathbf{A} }\ket{\B{p};0}|^2}{E-E'}
\end{equation}
where $E$ is the unperturbed energy of the initial state and primed quantities are those relevant to the intermediate state. Using the following relation for the dyadic Green's function (see, for example, \cite{Dung1998});
\begin{align}
\omega^2 \int d^3\B{r}'[\text{Im}\varepsilon(\B{r}',\omega)] {\B{G}}_{il}&(\B{r},\B{r}',\omega){\B{G}}^*_{lj}(\B{r}',\B{r}'',\omega) \notag  \\
&=\im {\B{G}}_{ij}(\B{r},\B{r}'',\omega)\label{DGFCompletenessResult}\; ,
\end{align}
 one finds for the energy shift \eqref{VacuumEnergyShift} in the no-recoil approximation (see \cite{Eberlein2004,Bennett2012})
\begin{equation}\label{GeneralShiftStartPoint}
\Delta E = \frac{ie^2}{2\pi m^2}{\langle p_i^2 \rangle } \int_{-\infty}^\infty \frac{d\omega}{\omega} \B{G}^\text{sc}_{ii}(\B{r},\B{r},\omega)\; ,
\end{equation}
where we have isolated surface-dependent effects by replacing $\B{G}$ with its scattering part $\B{G}^\text{sc}$ --- this is the portion of the Green's function that vanishes if all boundaries are removed. The Green's function $\B{G}^\text{sc}$ has no poles in the upper half of the complex-$\omega$ plane, so the entire shift integral can be worked out from its residue at $\omega=0$;
\begin{equation}\label{GeneralShift}
\Delta E = -\frac{e^2}{2 m^2}{\langle p_i^2 \rangle }\operatorname*{Res}_{\omega \to 0} \frac{\B{G}^\text{sc}_{ii}(\B{r},\B{r},\omega)}{\omega}\; .
\end{equation}
Equation \eqref{GeneralShift} is a new formula, containing all previous results for the surface-dependent self-energy (which can also be interpreted as a shift in mass) in specific situations (see \cite{Barton1988,Eberlein2004,Bennett2012}), but is valid for arbitrary surface geometry and material properties. To find the self-energy for a particular geometry one simply inserts the relevant scattering Green's tensor. For example, in the particular case of parallel perfectly conducting plates at $z=0$ and $z=d$ as considered in \cite{Barton1988}, the Green's tensor \cite{Tomas1995} expressed as a two-dimensional Fourier transform over parallel wave-vectors $k_\parallel$ is simple enough that the resulting $k_\parallel$ integral appearing in Eq.~\eqref{GeneralShift} can be carried out analytically. Defining $\zeta = z/d$ one finds after some algebra the self-energy shift between perfectly conducting plates;
\begin{align}\label{PCShift}
\Delta E_\text{PC}(\zeta) = &\frac{e^2\langle p_\parallel^2\rangle}{{32 m^2 \pi  d }} \Bigg\{ \pi\cot(\pi \zeta) \notag \\ &-2\left[\mathcal{H}_{3-\zeta } +\frac{3 \zeta^2-12 \zeta+11}{(\zeta-3) (\zeta-2) (\zeta-1)}\right]\Bigg\}  
\end{align}
where $\mathcal{H}_n \equiv  \sum_{k=1}^n k^{-1}$ and $p_\parallel$ is the momentum parallel to the plates. The above result agrees with \cite{Barton1988}.   Expanding for small $\zeta$ one finds that the leading term is given by $e^2 \langle p_\parallel^2\rangle/(32 \pi m^2 z )$, in agreement with previous single-plate work \cite{Eberlein2004,Bennett2012} obtained via a normal-mode quantization rather than macroscopic QED.

In realistic situations one requires more complex models of the surface response, for example the plasma model or Drude models defined respectively by 
\begin{equation}
\varepsilon_p(\omega) = 1-\frac{\omega_p^2}{\omega^2},\quad  \varepsilon_D(\omega) = 1-\frac{\omega_p^2}{\omega^2-\omega_\text{T}^2 +i \gamma \omega} \; . 
\end{equation}
In these situations, even the integrands of the $k_\parallel$ integrals become somewhat unwieldy so we do not report them here. It is interesting to note that the Drude model integral may again be carried out analytically, but the result contains hundreds of terms so is not particularly illuminating as compared to simply doing the integral numerically. Nevertheless, we agree with previous Drude model work for small $\zeta$ \cite{Bennett2012} and can still quote a reasonably compact new analytic result for the shift at $\zeta=1/2$ 
\begin{align}
&\Delta E^\text{mid}_{D}=-\frac{e^2 \langle p_\parallel^2 \rangle}{{256 \pi m^2  d^3 \EpsRatioStatic ^2}}\notag \\ 
&\times \Bigg\{\notag16 d^2 \EpsRatioStatic ^3 +8 \left[\EpsRatioStatic -1\right]^4  \ln [\EpsRatioStatic +1] \varepsilon'(0)^2 \notag \\
&+\bigg[ (\EpsRatioStatic -1)^2\left(2 \EpsRatioStatic  \varepsilon''(0)+\left(\EpsRatioStatic ^2-1\right) \varepsilon'(0)^2\right)\notag\\
&\times\!\! \Big(\!\EpsRatioStatic  \Phi \!\left[\EpsRatioStatic ^2,2,\nicefrac{1}{2}\right]\!+\!4 \text{Li}_2[\EpsRatioStatic]\!-\!3 \text{Li}_2\left[\EpsRatioStatic ^2\right] \! \Big)\! \bigg]\! \Bigg\}\label{FullDispersiveResultMidpoint}
\end{align}
where $\varepsilon'(0) \equiv [d\varepsilon(\omega)/d\omega ]|_{\omega=0}$, $\varepsilon''(0)\equiv [d^2\varepsilon(\omega)/d\omega^2] |_{\omega=0}$ and $\EpsRatio(\omega) \equiv  (\varepsilon(\omega)-1)/(\varepsilon(\omega)+1)$. We have made use of the polylogarithm function 
 $\text{Li}_s(x)\equiv \sum_{k=1}^\infty \frac{x^k}{k^s}$ and the `Lerch transcendent' $\Phi(x,s,\alpha)\equiv \sum_{k=0}^\infty \frac{x^k}{(k+\alpha)^s}$. 
 
\begin{figure}[h!] 
\includegraphics[width = 0.85\columnwidth]{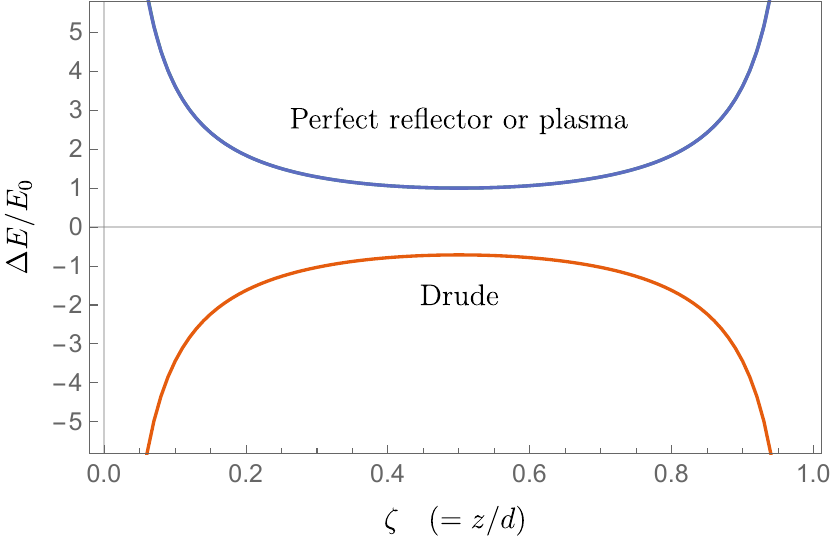}
\caption{Surface-dependent self-energy for Drude and plasma models, as a function of position $z$ between plates separated by a distance $d$. Here we have scaled the energy by the energy shift $E_0$ of the electron at the midpoint between parallel perfect conductors, obtained from Eq.~\eqref{PCShift} by setting $\zeta=1/2$ with result (in S.I. units) $E_0 = e^2 p_\parallel^2 \ln 2 /(8\pi \epsilon_0 c^2 m^2 d)$. This is approximately equal to $8 \times 10^{-28}$J ($\sim 5 $neV) for an electron moving at $0.01$c in a $10\mu$m-wide gold cavity ($\omega_p = 1.37 \times 10^{16} \text{Hz},\gamma = 4.05 \times 10^{13} \text{Hz}$ and $\omega_\text{T} \approx 10^{15} \text{Hz}$ \cite{Ordal1985})}
\label{DrudePlasmaGraph}
\end{figure}

 We plot the $\zeta$ dependence of the shift in Fig.~\ref{DrudePlasmaGraph},
where we see the unexpected effect that the energy shift has a different sign the for the Drude and plasma models. 

  An intuitive explanation could be formed by considering  this energy shift in the context of mass
renormalization: the coupling of the electron to the electromagnetic field causes an increase from some
fixed ÔbareÕ mass $m_0$ to the mass m that we observe, with $m = m_0 + \delta m$, where $\delta m$ includes the mass shift in free
space as well as any surface dependent correction such as that calculated here. The mass shift $\delta m$ of a free particle is related to the energy shift calculated here via $\frac{\delta m}{2m^2}\langle \B{p}^2\rangle  = -\Delta E$ (cf. \cite{milonni1994quantum}), so that a positive surface-dependent component in the energy shift corresponds to a negative surface-dependent correction to $\delta m$. Thus our plasma results, for example, correspond to a small reduction in mass compared to that in free space. This makes sense in terms of the fact that our result is dominated by low-frequency excitations, where the plasma model behaves very similarly to a perfect conductor; an incident electric field is completely reflected and undergoes a $\pi$ phase shift.  This causes it to destructively interfere with the incident wave \footnote{The integrals are dominated by small $k_\parallel$ due to their exponential falloff, meaning it suffices to consider normal incidence.}. This means that the electron feels a weaker effect from radiation reaction than it would do if it were in free space, meaning that the surface-dependent contribution to the mass is negative, as borne out in the results presented here.  Conversely, the Drude surface behaves more like a dielectric at low frequencies, with polariton excitations opening
up additional channels for the interaction with the electron, resulting in an increase in the mass relative to that for free space.

This sign difference between Drude and plasma models is the kind of feature that is relatively easy to measure experimentally, so in the following sections we explore a possible method for experimental diagnosis of wether Drude or plasma models are more appropriate for quantum field theory near a given surface.

\begin{figure}[h!]
\includegraphics[width = 0.85\columnwidth]{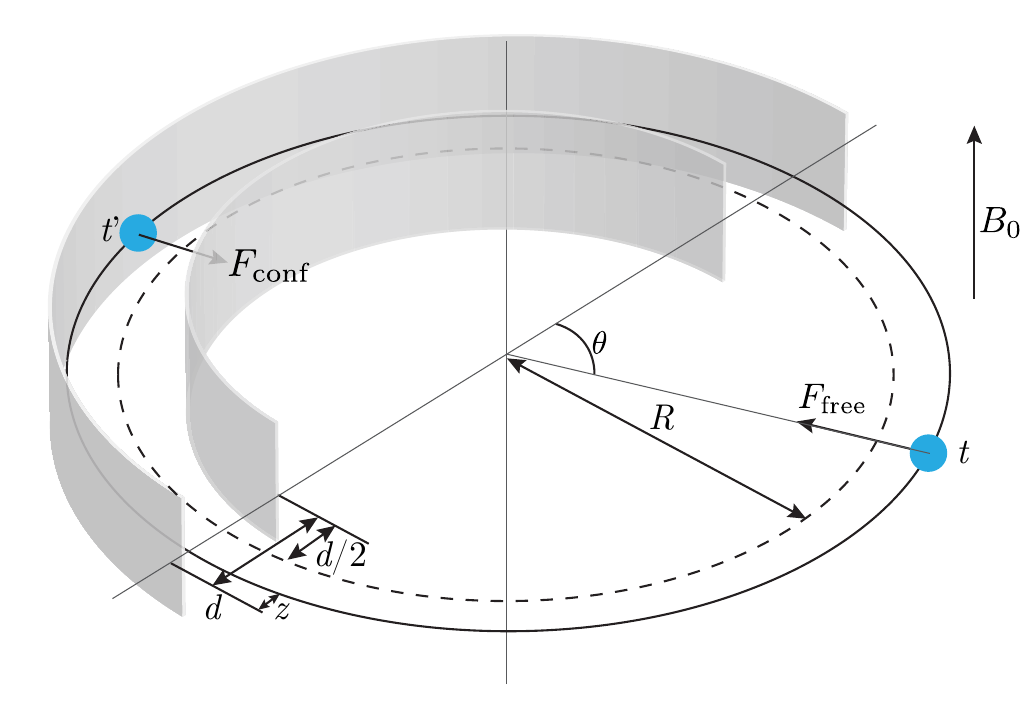}
\caption{Schematic of the proposed `half-cyclotron' experimental setup, where half of the electron's cyclotron orbit is subject to a surface-dependent correction and half is not.}
\label{HalfRing}
\end{figure}

An experiment aiming to measure the shift discussed above must involve an electron that can be moved in and out of proximity to a surface, and ideally it should also be set up in a way that the experimental observable depends as strongly as possible on the sign of the dynamical force found from \eqref{GeneralShift}, which of course is an addition to the ever-present (and much larger, at least in a non-relativistic setting) electrostatic force. For example, sending an electron beam through a cavity in an experiment analogous to the Casimir-Polder force experiment of \cite{Sukenik1993} and attempting to observe the momentum-dependence of the deflection would be extremely difficult as the Drude/plasma difference in the $\mathcal{O}(\nicefrac{v}{c})^2$ dynamical force would show up as an unobservably tiny additional deflection to that given by the purely electrostatic (velocity-independent) force \footnote{E.g. an electron $1\mu$m away from a conductor is attracted to its image and collides with the surface in about $0.1$ns, setting an upper bound on the time for which the electron can be used. If it moves parallel to the surface at $0.1$c, the dynamical force calculated here causes an additional deflection of the order $1$nm.}. Thus a radically different approach is required, for which we propose the setup shown in Fig.~\ref{HalfRing}.
Here, an electron undergoes cyclotron motion in a tube extending halfway around a circle (a `half-cyclotron') and ultimately we will show that with reasonable parameters the Drude and plasma models could be distinguished by observing what type of magnetic field modulation is required to keep the electron in persistent cyclotron motion.

We describe the apparatus by considering the electron to be `free' (aside from the magnetic field causing cyclotron motion) when in the region $0<\theta<\pi$, while in the region $\pi < \theta < 2\pi$ it is considered as confined in the tube. In this first proof-of-principle calculation we take $R \gg d$ and hence ignore edge effects in the transition regions $\theta \approx 0$ and $\theta \approx \pi$ (though in principle these could be estimated from known exact Green's functions for systems with edges \cite{Eberlein2007,*Eberlein2009}), giving for the force acting on the electron:
\begin{equation}
F(\theta) = \begin{cases}
F_\text{free} & \text{for} \quad 0 < \theta \leq \pi\\
F_\text{conf} & \text{for} \quad \pi <\theta \leq  2\pi 
\end{cases} \; . 
\end{equation}
The assumption $R\gg d$ means that the curved section can be considered locally as parallel plates, so the force $F_\text{conf}$ up to order $\langle p_\parallel^2  \rangle $ can be obtained from the results of the previous section. Thus we have a prescription that allows us to consider the forces upon an electron that moves around the complete circle. The acceleration of the electron should have a negligible effect on the electron-plate interaction itself as the parameters we will choose are well within the regime where the acceleration is much less than $c^2/z$ \cite{Marino2014a}.

 Considering first the free region, the force on the circulating electron is given simply by the cyclotron expression;
$
F_\text{free} = e v B_0
$
with $B_0$ as appropriate to cause the electron's cyclotron radius to match the radius of curvature of the tube. The electron then passes through the confined region, where, assuming it is not exactly in the center of the tube, it gains an additional surface dependence. This will cause the electron to leave its circular orbit meaning that the applied $B_0$ is no longer appropriate for cyclotron motion.  In order to remedy this, one has to change the magnetic field by an amount $\Delta B$ in such a way that the effective force on the electron (when it is in the confined region) is always $evB_0$, i.e. we choose $\Delta B $ so that $e v (B_0+\Delta B)+ F_\text{surf}= e v B_0$ is satisfied, meaning that the required field modulation is $\Delta B = -{F_\text{surf}}/(e\beta c )$ where  $\beta \equiv v/c$. 

\begin{figure}[t!]
\includegraphics[width = \columnwidth]{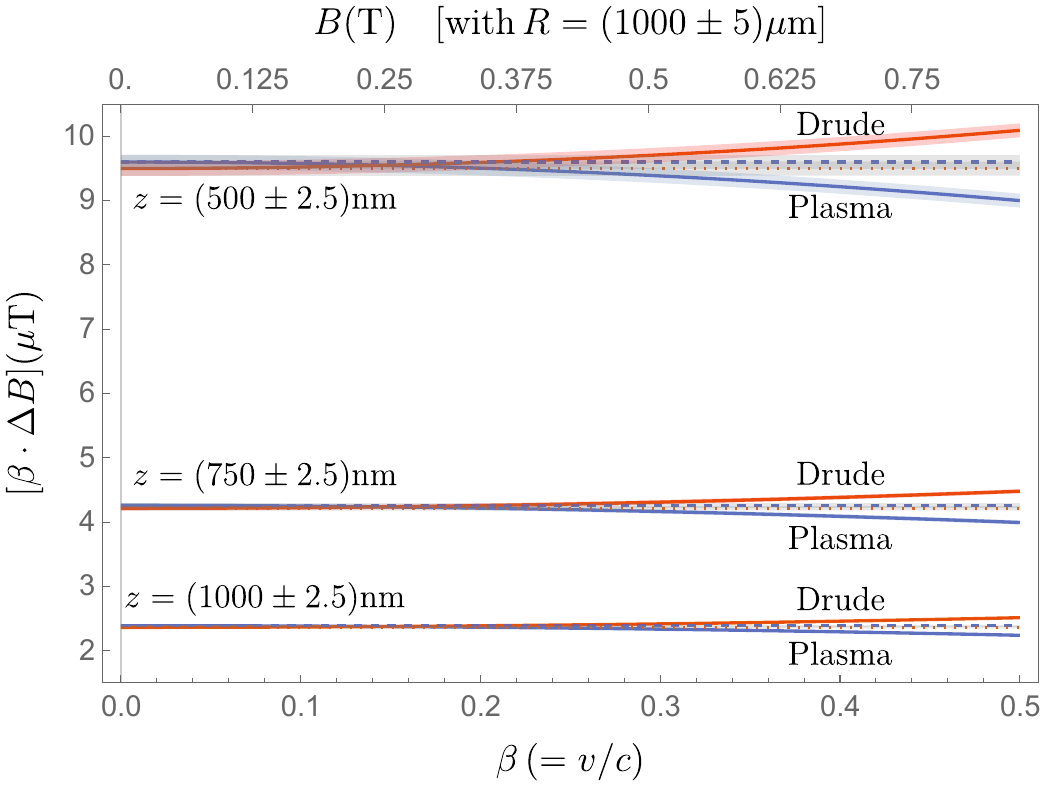}
\caption{Field modulation strength required to keep an electron with velocity $v=\beta$c in cyclotron motion in the apparatus shown in Fig.~\ref{HalfRing}, and the various positions $z$ for the electron are taken as indicated in the figure, each with an uncertainty of $\Delta z = 2.5$nm associated with a narrow electron beam \cite{Fischbein2008}. The stability of the magnetic field is taken as one part in $10^5$, as discussed in the main text. We also include a $5\mu$m variance in the radius of the ring and a $1\mu$m variance in $d$ to account for possible manufacturing imperfections --- the former turns out to be by far the dominant error.  The solid lines (blue for plasma, red for Drude) represent the results with all dynamical effects included, while the dashed (Drude) and dotted (plasma) lines are those when only the static terms are taken. The red (blue) shaded areas are the estimated uncertainties for the dynamical shifts required for Drude (plasma) models, while the grey shaded areas are the uncertainties in the static fields.  Since the electron is undergoing cyclotron motion, each velocity implies a particular magnetic field strength (for a given radius), given simply via $B = (mc/e) \beta/R \approx (0.0017\text{ T}\cdotp \text{m})\beta/R$ --- this magnetic field (for $R=1$mm) is indicated on the upper axis.}\label{TripletGraph}
\end{figure}

The surface force $F_\text{surf}$ consists of an electrostatic part, as well as the first dynamical correction calculated in the previous sections. The electrostatic part can be found via textbook calculation and is given by
\begin{equation}
F_\text{static} = \frac{\eta(0)  e^2}{16 \pi  d^2 }\left[\Phi \left(\eta(0)  ^2,2,1-\zeta\right)-\Phi \left(\eta(0)  ^2,2,\zeta\right)\right]
\end{equation}
Initially taking into account just the electrostatic part of the force, one has $\Delta B_\text{static} =-F_\text{static}/(e\beta c)$  for the required field modulation. The static force is of course independent of $\beta$, so the quantity $\beta \Delta B_\text{static}$ is constant in $\beta$, as demonstrated in Fig.~\ref{TripletGraph}. 
There we have chosen parameters such that the effects of surface roughness and patch charges should be minimised. For gold surfaces the RMS roughness and patch size can be as low as 0.4nm \cite{Ederth2000} and 25nm \cite{Decca2005a} respectively, so all the electron-surface distances chosen are orders of magnitude greater than these length scales. While a full analysis of roughness and patch effects is far beyond the scope of this proof-of-principle work, previous investigations of the corresponding corrections to the Casimir force show that both these effects are negligible when plate separation significantly exceeds roughness amplitude  \cite{VanZwol2008} and patch size \cite{Decca2005a, Behunin2012}.

\begin{figure}[t!]
\includegraphics[width = \columnwidth]{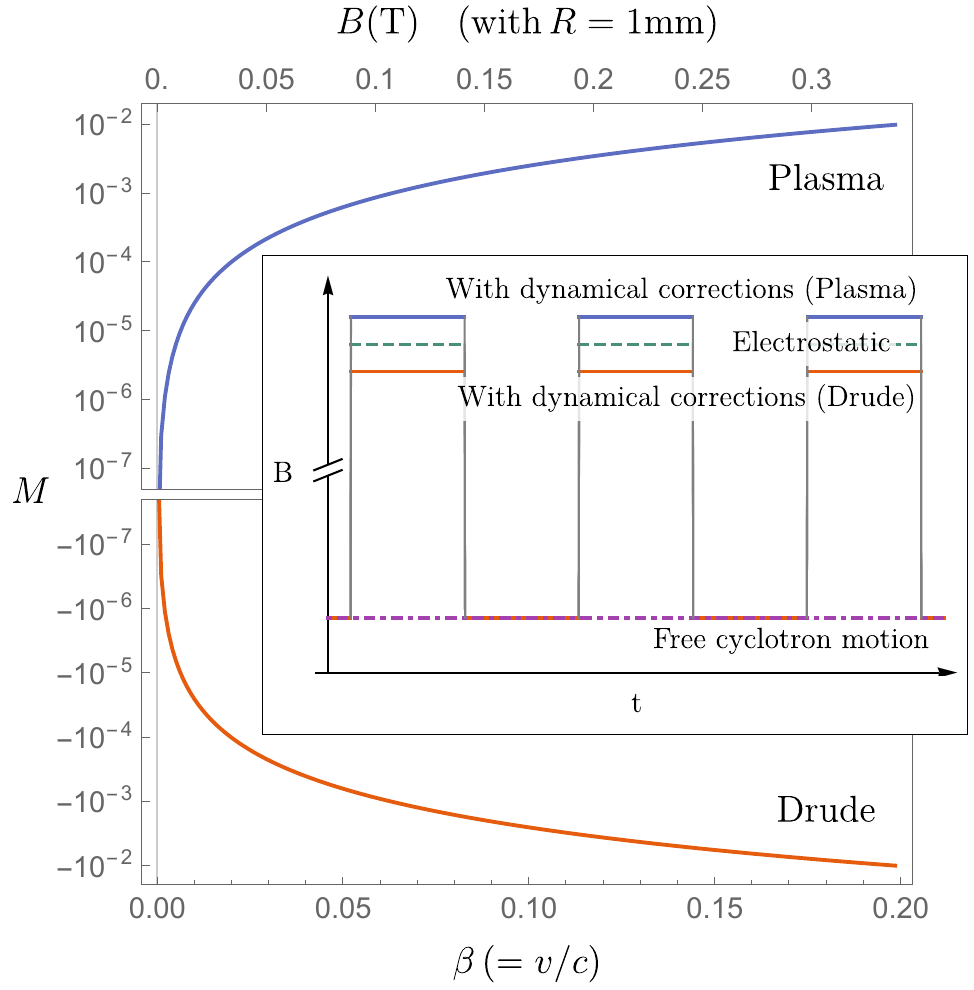}
\caption{Relative field correction factor $M$ needed in order to preserve cyclotron motion when dynamical surface effects are included at $z=1\mu$m. If no dynamical surface effects were present, one would have $M=0$. The experimental errors are taken as the same as in fig.~\ref{TripletGraph}, which turn out to be too small to be visible at the scale of this graph. Inset: Schematic representation of the time dependence of the $B$-field required to preserve cyclotron motion for both models.}\label{RRM}
\end{figure}

If we now derive a force $F_\text{dyn}$ from the dynamical energy shift $\Delta E$ via $F_\text{dyn}=-d\Delta E/dz$, we can then add this to the electrostatic part to find the required field modulation when dynamical corrections are taken into account;
\begin{equation} \label{DeltaBFullDefinition}
\Delta B=  -\frac{F_\text{static}+F_\text{dyn}(\beta)}{e\beta c },
\end{equation}
Now the product $\beta \Delta B$ will no longer be constant in $\beta$, as demonstrated by the solid lines in Fig.~\ref{TripletGraph}. It is important to note that Fig.~\ref{TripletGraph} pushes the bounds of our model (large velocities) and of experimental reality (small distances) in order to demonstrate a general trend. For the more realistic situation of smaller velocities and larger distances, it is more convenient to investigate the dimensionless quantity $M$, defined as; 
\begin{equation}
M\equiv -\frac{F_\text{dyn}}{F_\text{static}}=\frac{\Delta B}{\Delta B_\text{static}}-1
\end{equation}
where the equality follows from Eq.~\eqref{DeltaBFullDefinition} and the definition of $\Delta B_\text{static}$. This quantity is a measure of how large the field modulation that preserves cyclotron motion needs to be if all dynamical corrections are included, relative to that required if there were no dynamical effects.  For perfect reflectors, the small $\zeta$ approximation of $M$ is particularly simple:
$
M_\text{PM}(\zeta \approx 0) = \frac{\beta^2}{4} + \mathcal{O}(\zeta ^3),
$
while for realistic models we plot $|M|$ in Fig.~\ref{RRM}.
For example, given a magnetic field of $0.035\text{T} = 350\text{G}$ one would have to add or subtract a modulation with a relative magnitude of $10^{-4}$, which is $100$ parts per million (ppm). Magnetic field sources routinely have stability at the $0.1-5$ppm scale \cite{Carey1999,Roberts2001} (which compares favourably with the 10ppm taken in fig.~\ref{TripletGraph}), and in extreme cases can approach one part per billion \cite{Odom2006}. This means that with routinely-achievable magnetic field stability the required modulation would be approximately three orders of magnitude larger than the background magnetic field instability.

In this work we have derived a general formula for the shift in the self-energy of an electron in arbitrary environments. We have applied this to the situation of identical parallel plates, reproducing in the relevant regime earlier results obtained by normal-mode quantization near a single plate. We found an unexpected result whereby the dynamical shift for Drude and plasma models of the surface are of different signs, and approximately the same magnitude. We then outlined a cyclotron motion-based experiment that takes advantage of this specific feature of our results in order to distinguish whether a Drude or plasma model is more appropriate for a given surface. This setup proposed here would provide a reliable, independent, and experimentally clean probe of macroscopic media's low-frequency response to the fluctuating vacuum, which has been at the heart of continuing issues in Casimir physics. It is not clear whether the root of this debate is indeed a fundamental issue or a hitherto unresolved problem with experimental tests, but having an additional and independent method to shed more and most importantly new light on the issue is certainly a step into the right direction towards resolution, in one way or another.
\begin{acknowledgments}
This work was supported by the Deutsche Forschungsgemeinschaft (grants BU 1803/3-1476), the Alexander von Humboldt Foundation, the UK Engineering and Physical Sciences Research Council (EPSRC) and the Freiburg Institute for Advanced Studies.
\end{acknowledgments}

%

\end{document}